# Kill Webs by Collaborative & Self-organizing Agents (CSOAs)

Ying Zhao

*Naval Postgraduate School*

Monterey, CA 93943, USA

yzhao@nps.edu (ORCID:0000-0001-8350-4033)

Charles C. Zhou

*Quantum Intelligence, Inc.*

Prunedale, CA 93907

charles.zhou@quantumii.com (ORCID: 0000-0001-9598-015X)

*Abstract*—A single agent represents a single system capable of ingesting local data, indexing, cataloging information, performing knowledge pattern discovery, and separating patterns and anomalies from data. Multiple agents work collaboratively in a peer-to-peer network. Each agent has a peer list. Such multiple agents' collaboration can be modeled as cooperative games. Each agent optimizes its own objective locally. We show that each agent self-organizes or converges to its best value and the whole agent network achieves the best social welfare based on both the quantum adiabatic evolution transformation (QAET), and quantum intelligence game (QIG) or the QAET-QIG framework. We apply the QAET-QIG framework to the kill web concept that can potentially improve the traditional kill chain process or the find, fix, track, target, engage, and assess (F2T2EA) process. The improvement is measured in the values of powerful global optimization, distributed lethality, and load balancing. We show a use case of the QAET-QIG frame in a potential application of mixed sensors, platforms, weapons, and effects.

## I. INTRODUCTION

A single agent [1], [2] ingests local data and performs data mining and machine learning. Multiple agents work collaboratively in a peer-to-peer network via a peer list. When agents learn in an unsupervised fashion, they are collaborative and self-organizing agents (CSOAs). We consider the CSOAs' interaction as an evolution process of quantum superpositions.

### A. Quantum Evolution

A quantum evolution is modeled as a time-dependent quantum evolution characterized by Hamiltonians in Schrödinger's equations. Collaborative and self-organizing behavior of CSOAs in the process result in a total value of an agent with respect to its peer network. Equation (1) is the time-dependent Schrödinger equation and evolution of a quantum state $\psi(\vec{x}, t)$, where $\vec{x}$ is a space vector representing other variables that might depend on $t$ and is omitted for simplicity, i.e, $\psi(t) = \psi(\vec{x}, t)$:

$$i\hbar \frac{d}{dt} \psi(\vec{x}, t) = H(t) \psi(\vec{x}, t). \quad (1)$$

Equation (2) shows the solution to Equation (1):

$$\psi(\vec{x}, t) = e^{-iH(t)t/\hbar} \psi(\vec{x}, 0) = U(t) \psi(\vec{x}, 0). \quad (2)$$

The operator $U(t) = e^{-iH(t)t/\hbar}$ is the time-evolution operator and unitary. The eigenstates of the Hamiltonian $H(t)$, known as energy eigenstates, are the solutions to the Schrödinger equation, form a complete basis set for the state space of the quantum system. A density matrix $\rho$ of the wave function $\psi(t)$ is governed by the underlying $H(t)$. The evolution of $\rho$ over time is determined by the energy eigenstates.

Such an unsupervised learning system optimizes its value (e.g., energy or entropy) using natural mechanisms such as quantum adiabatic evolution transformation (QAET) [6]. According to the adiabatic theory, a QAET refers to a slow, continuous change of the Hamiltonian of a quantum system, where if the change is sufficiently slow or adiabatic, the system remains in its instantaneous ground state throughout the evolution.

A QAET can play an essential role to generate more pure and entangled quantum states that are useful such as creating robust topological orders for condensed materials [7] and semantic innovation in finance [8]. In this paper, we focus on the application of these concepts to kill webs.

In a traditional QAET setting [6], one assumes the beginning Hamiltonian is $H_B$ and the ending Hamiltonian is $H_C$, the objective is to optimize the final measurement of $|\psi(T, \gamma, \beta)\rangle$ at time $T$ by changing $\gamma = (\gamma_1, ..., \gamma_p)$ and $\beta = (\beta_1, ..., \beta_p)$ in Equation (3) and Fig. 1 using hybrid classic and quantum computing to optimize parameters $\beta_t$ and $\gamma_t$.

$$|\psi(T, \gamma, \beta)\rangle = \prod_{t=1}^{T} U_B(\beta_t) U_C(\gamma_t) |\psi(0)\rangle, \quad (3)$$

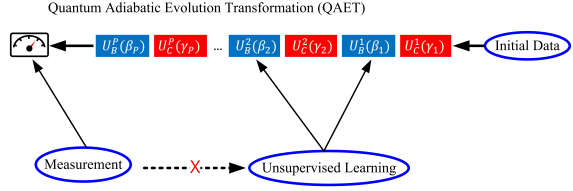

Fig. 1: Traditional quantum adiabatic evolution transformation (QAET): QAET needs a series of quantum unitary transformations using hybrid classic and quantum computing to optimize coherence parameters $\beta_p$ and $\gamma_p$.

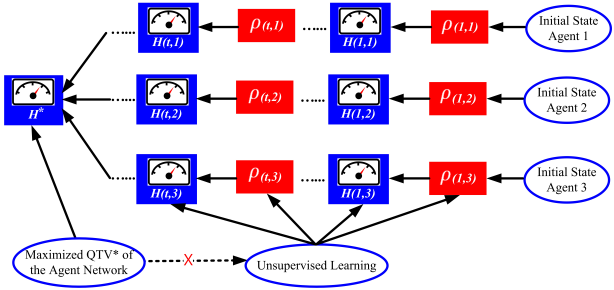

Fig. 2: Multi-Agent Collaboration. QAET-QIG : The value of a multi-agent system increases in a natural quantum mechanism QAET with repeated measurements in a QIG. The two processes are combined.

where

$$U_C(\gamma_t) = e^{-i\gamma_t H_C} \quad (4)$$
$$U_B(\beta_t) = e^{-i\beta_t H_B}. \quad (5)$$

To compute $\gamma_t, \beta_t$ in the traditional QAET, a QAET process slowly changes the system's Hamiltonian from $H_B$ to $H_C$ by designing a trajectory $s(t)$ to obtain the maximum energy state of $H_C$ using Equation (6) [21]:

$$H(t) = [1 - s(t)]H_B + s(t)H_C \quad (6)$$

where $s(t)$ is a smooth function, $0 \leq s(t) \leq 1$, $s(0) = 0$ and $s(T) = 1$.

In reality, external environment can change the Hamiltonian of a quantum system, for example, a measurement alters the system state based on the previous $H_{t-1}$'s eigenstate, collapses the system's wave function into the ground eigenstate. We show that such a measurement combined with a quantum intelligence game (QIG) forced by the environment can be an alternative method to the traditional QAET of changing $H_B$ to $H_C$.

### B. Link Knowledge Graphs and Kill Webs

The knowledge graphs are embedded in the environment measurement of the Hamiltonian $H$. A team of CSOAs carry out a task with the following assumptions:

- The environment is modeled as the time-dependent Hamiltonian of a multi-agent system.
- When more capabilities are used together, the corresponding $H$ graph might be directed, asymmetric, and causal, in other words, $H$ might be non-Hermitian, which is different from the traditional quantum theory Hermitian matrix.
- Each agent does not have a full knowledge of the entire Hamiltonian and has to estimate from data-driven approaches learning from the environment shown in Fig. 2.

Such a graph-structured Hamiltonian measures causal relations between subcomponents and defines quantum states, linking to active research area of quantum causal models and quantum Bayesian networks. The contributions of this paper are summarized as follows:

1) We show that a multi-agent quantum system converges to an equilibrium state which has a high degree of purity and coherence via a QAET-QIG evolution in a superpositioned torus. The convergence results in an equilibrium and measurable property or quantum theoretic value (QTV) of the system that is optimized. We demonstrate the QAET-QIG framework in Fig. 2, show two theorems, an algorithm used to score new data for quantum values, and compare them with the traditional QAET (Fig. 1).
2) We apply the QAET-QIG framework to the kill web concept that potentially improve the traditional kill chain process such as find, fix, track, target, engage, and assess (F2T2EA). The improvement is measured in the values of powerful global optimization, distributed lethality, and load balancing. We show a use case of the QAET-QIG framework to a potential application of mixed sensors, platforms, weapons, and effects.

## II. RELATED METHODS

### A. Quantum Properties

In quantum mechanics, the state of a quantum system is described by a wave function $\psi$ or, more generally, by a density matrix $\rho$. A density matrix is more insightful to describe if a quantum state is pure or mixed, if there are quantum coherence between states. Consider a multi-agent system behaves in a quantum sense, various quantum properties [10] can be computed.

- Purity: A pure state is a state that can be described by a single wave function. Equation (7) shows the purity calculation:

$$Purity = \sum_{k=0}^{K} \lambda_k^2, \quad (7)$$

where $\lambda_k$ is the eigenvalues of the density matrix of a quantum system.

- Quantum entanglement entropy (QEE): QEE characterizes the randomness or disorder within a multi-agent system $\rho$. The eigenvalues $\lambda_k, k = 0, ..., K$ of $\rho$, represent probabilities, so $\lambda_k > 0, k = 0, ..., K$ and $\sum_{k=0}^{K} \lambda_k = 1$.

$$QEE = -\sum_{k=0}^{K} \lambda_k ln(\lambda_k) \quad (8)$$

$QEE$ for the whole system in a pure state (Purity = 1) is 0 according to Equation (8). QEE can be used to measure quantum topological phase transitions [14] or so called quantum topological order [5].
- Coherence: Coherence refers to the phases of quantum probability amplitudes and enables quantum superposition and interference. Coherence is measured using the $L_1$ norm of the off-diagonal elements of $\rho$ as shown in Equation (9).

$$Coherence = \sum_{k \neq l} |\rho_{kl}| \quad (9)$$

### B. Application of Perron-Frobenius Theorem and QAET

The Perron-Frobenius theorem [16] is primarily concerning non-negative matrices (i.e., matrices with all non-negative entries), or more specifically, irreducible non-negative matrices. Such a matrix has a unique largest positive eigenvalue, which is also the spectral radius of the matrix, the corresponding eigenvector has all its components positive for the largest eigenvalue. The eigenvalue is strictly larger in magnitude than all other eigenvalues. The Perron-Frobenius theorem provides the mathematical reasoning that a ground state exists when a wave function "collapses." If the operator of a measurement $H$ is nonnegative and irreducible, the real eigenvalue corresponds to a state that remains physically measurable, while the other states corresponding to complex eigenvalues do not.

### C. Game Theory: An Unsupervised and Self-Organizing Mechanism

The unsupervised and self-organizing mechanism, QAET-QIG, is achieved by an application of the Nash equilibrium (NE) [18]. In a classic or quantum game, NE characterizes a strategy $\psi_l^*$ discourages a unilateral deviation such that

$$u_l(\psi_l^*) \geq u_l(\psi_l, \psi_{-l}^*) \quad (10)$$

for all $\psi_l$ and $l$. $(\psi_l, \psi_{-l})$ is the choice of player $l$ (self-player) relative to all other players $-l$ (opponent). $u$ is a utility or measurement function.

## III. RESULTS

This section we present the main theorem of the paper.

**Theorem 1** (QAET-QIG Evolves to a quantum theoretic value (QTV)). *A multi-agent system, with a Hamiltonian $H$ environment measurement, which is assumed non-negative and irreducible, converges to an equilibrium state via a modified QAET process or a QAET-QIG process shown in Fig. 2. The evolution results in an equilibrium measurement. QTV of the system is the equilibrium measurement or value in Equation (11) when $t \to \infty$. Each individual agent also reaches its own optimal value.*

$$QTV(t+1) = \langle \psi(t) | H(t+1) | \psi(t) \rangle \quad (11)$$

**Proof.** Consider a wave function of superposition in Equation (12), $\hbar$ is adsorbed into $E_k$:

$$|\psi(t)\rangle = \sum_{k=0}^{K} \sqrt{\lambda_k} e^{-iE_k t} |\psi_k\rangle, \quad (12)$$

where $E_k$ and $|\psi_k\rangle, k = 0, ..., K$ are instantaneous eigenvalues and eigenvectors of $H(t)$, respectively. $|\psi_k\rangle$ can be functions of other attributes than time $t$, e.g., 3D location X, Y, Z, omitted here for the proof, however, we give an example later showing this case. $\vec{\lambda} = \begin{bmatrix} \lambda_0 \\ ... \\ \lambda_K \end{bmatrix}$, $\sum_{k=0}^{K} \lambda_k = 1$, and $\lambda_k > 0$ represent the probability amplitude of the superposition for time $t$. The density matrix is represented in Equation (13).

$$\rho(t) = |\psi(t)\rangle \langle \psi(t)|$$
$$= \begin{bmatrix} \lambda_0 & \sqrt{\lambda_0 \lambda_1} e^{-i(E_0-E_1)t/\hbar} & ... \\ \sqrt{\lambda_1 \lambda_0} e^{-i(E_1-E_0)t/\hbar} & \lambda_1 & ... \\ ... & & \lambda_K \end{bmatrix} \quad (13)$$

$\rho(t)$ is Hermitian: $\rho(t)^\dagger = \rho(t)$; idempotent: $\rho(t)^2 = (|\psi(t)\rangle\langle\psi(t)|)^2 = \rho(t)$, and Trace one: $\text{Tr}(\rho(t)) = \langle\psi(t)|\psi(t)\rangle = \sum_{k=0}^{K} \lambda_k = 1$. These properties imply that $\rho$ has exactly one eigenvalue equal to 1, and all other eigenvalues are 0. So $\rho(t)$ is a rank-1 matrix and pure although it has multiple nonzero entries. In other words, $\psi(t)$ constructed in Equation (12) of a pure state of a complete coherent superposition.

$QTV(t+1)$ can be computed recursively. $|\psi(t)\rangle$ is measured on $H(t+1)$ where

$$H(t+1) = (V(t+1))^\dagger E(t+1) V(t+1) \quad (14)$$

is the eigenvalue decomposition of $H(t+1)$ with eigenvalues $E(t+1)$ and eigenstates $V(t+1)$, when H is a Hermitian. Let

$$|\psi'(t)\rangle = V(t+1) |\psi(t)\rangle, \quad (15)$$

Assume $H(t+1)$ is non-negative and irreducible, based on the Perron-Frobenius theorem, there is an unique real eigenvalue with the maximum magnitude, i.e., $|E_0(t+1)|$. $|E_0(t+1)| \geq |E_1(t+1)| \geq ... \geq |E_K(t+1)| \geq 0$.

$$\begin{aligned} QTV(t+1) &= \\ &= \langle \psi(t) | H(t+1) | \psi(t) \rangle \\ &= \langle \psi(t) | (V(t+1))^\dagger E(t+1) V(t+1) | \psi(t) \rangle \\ &= \sum_{k=0}^{K} |E_k(t+1)| \langle \psi'_k(t) | \psi'_k(t) \rangle \\ &= \sum_{k=0}^{K} |E_k(t+1)| \lambda_k(t) \\ &\leq |E_0(t+1)| \end{aligned} \quad (16)$$

Equation (16) shows $QTV(t+1)$ is a weighted sum of eigenvalues of $H(t+1)$ and the weights $\vec{\lambda}(t) = \begin{bmatrix} \lambda_0(t) \\ ... \\ \lambda_K(t) \end{bmatrix}$ are the probability amplitudes of the quantum state $|\psi(t)\rangle$ measured at the eigenstates $V(t+1)$ of $H(t+1)$.

In the QAET-QIG mechanism, in addition to the superposition Equation (12) that is critical, the quantum effect is further reinforced by using a QIG. We assume each quantum agent $l$ tries to change the quantum configuration component of itself $\vec{\lambda}$ to increase its own value. Each agent's value is $\lambda_l$. According to the notation in Equation (10), $\vec{\lambda}$ is the quantum superposition state in Equation (12): Given the rest of the agents' state is $\psi_{-l}(\vec{\lambda})$ and value is $\lambda_{-l}(\vec{\lambda})$, $\psi_l$ can be only $\psi_l(\vec{\lambda})$ because the parts of a pure state are not independent, they "move" or behave like a single, inseparable whole.

$$\lambda_l + \lambda_{-l} = 1. \tag{17}$$

In realistic quantum processes, the $H$ is often not known in advance, it acts on the quantum state $|\psi\rangle$, the system undergoes a collapse with a probability $\lambda_l$. This means that components of $|\psi\rangle$ which better align with the dominant eigenvectors of $H$ are more likely to survive the measurement.

∎

Consider $K = 1$ in Equation (12), $\psi_0$ and $\psi_1$ are two orthogonal circles for data attributes such as locations (X,Y, Z). Super-positioning the $\psi_0$ and $\psi_1$.

$$\psi(X,Y,Z,t) = \sqrt{\lambda_0}e^{-iE_0 t}\psi_0(X,Y) + \sqrt{\lambda_1}e^{-iE_1 t}\psi_1(Y,Z) \tag{18}$$

represents a quantum point in the three dimensional space evolves according to Equation (19), a torus in Fig. 3 as a result of a superposition and evolution of two eigen-states. Fig. 4 shows the probability amplitude in the torus.

$$\begin{aligned} X &= (R + r\cos(\theta(t)))\cos(\phi(t)) \\ Y &= (R + r\cos(\theta(t)))\sin(\phi(t)) \\ Z &= r\sin(\theta(t)) \end{aligned} \tag{19}$$

where

$$\begin{aligned} R &= \lambda_0 + \lambda_1 = 1 \\ r &= 2\sqrt{\lambda_0 \lambda_1} \\ \theta(t) &= E_0 t \quad \phi(t) = E_1 t \end{aligned}$$

### A. Scoring a Quantum State on Rank-1 Measurement H

Given a quantum state constrained to the torus-evolving form:

$$\psi(t) = \sqrt{\lambda_0}e^{-iE_0 t}\psi_0 + \sqrt{\lambda_1}e^{-iE_1 t}\psi_1, \tag{20}$$

where $\psi_0, \psi_1$ are orthonormal, and $\lambda_0 + \lambda_1 = 1$. We are considering a parametric measurement operator $H$:

$$H(\alpha) = \begin{bmatrix} \cos^2\alpha & \cos\alpha\sin\alpha \\ \cos\alpha\sin\alpha & \sin^2\alpha \end{bmatrix}. \tag{21}$$

This is a rank-1 matrix since the maximum eigenvalue is 1. The QTV is maximized when

$$\begin{bmatrix} \sqrt{\lambda_0} \\ \sqrt{\lambda_1} \end{bmatrix} = \begin{bmatrix} \cos\alpha \\ \sin\alpha \end{bmatrix}, \tag{22}$$

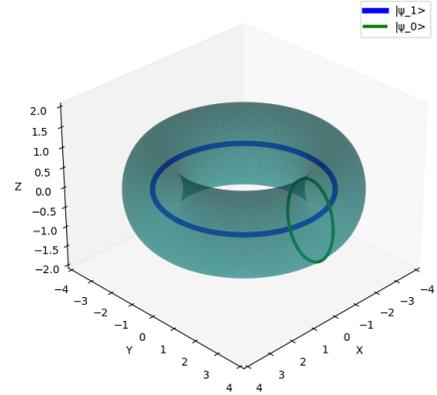

Fig. 3: A Torus as a Result of a Superposition and Evolution of Two Eigenstates

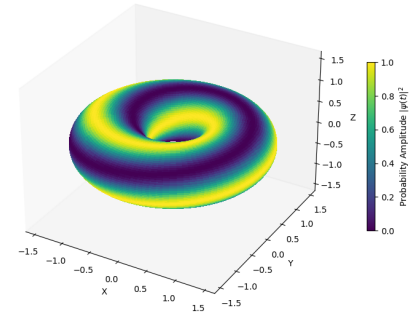

Fig. 4: Probability Amplitudes of a Quantum State on the Torus Formed by Super-positioning Two Orthogonal Eigenstates as Bases

and

$$t = \frac{2n\pi}{E_1 - E_0}, \tag{23}$$

where $n$ is an integer. In other words, when the state aligns the best to the top eigen vector of the measurement $H$, we have

$$\begin{aligned} QTV_{max\_rank-1} &= \max_t \langle\psi(t)|H(\alpha)|\psi(t)\rangle \\ &= (\cos\alpha\sqrt{(1-\lambda_1)} + \sin\alpha\sqrt{\lambda_1})^2 \\ &= 1 \end{aligned} \tag{24}$$

### B. Classical State with Quantum Measurement

Let the system be described by an incoherent mixture (a classical probability distribution over orthogonal states):

$$\rho = (1-\lambda_1)|\psi_0\rangle\langle\psi_0| + \lambda_1|\psi_1\rangle\langle\psi_1|. \tag{25}$$

Now consider the same measurement $H$ with off-diagonal components:

$$H = \begin{bmatrix} a & c \\ d & b \end{bmatrix} \quad \text{in basis } \{\psi_0, \psi_1\}. \tag{26}$$

The expectation value of the measurement such a quantum system is:

$$QTV_{max\_mixed} = \langle H \rangle = \text{Tr}(\rho H) = (1 - \lambda_1)a + \lambda_1 b < 1 \tag{27}$$

### C. Superposition Implies a Torus Evolution

The quantum nature is that a quantum state viewed as a distribution over a torus. This form of $\rho$ contains off-diagonal coherence:

$$\rho_{01}(t) \sim \sqrt{\lambda_0 \lambda_1}\, e^{i(E_1 - E_0)t}. \tag{28}$$

When measured with an observable $H$ that has off-diagonal elements as well, the measurement manifests the quantum coherence when $e^{i(E_1 - E_0)t} = 1$ and $t$ is shown in Equation (23).

**Theorem 2.** *For any quantum state that is a coherent superposition of two orthonormal energy eigenstates $\psi_0$ and $\psi_1$, or pure states in $\mathcal{H}_A \otimes \mathcal{H}_B$ with distinct eigenvalues $E_0, E_1$,*

$$\psi(t) = \sqrt{\lambda_0}\, e^{-iE_0 t}\psi_0 + \sqrt{\lambda_1}\, e^{-iE_1 t}\psi_1,$$

*the time-evolved state traces a torus $\mathbb{T}^2 \subset Hilbert\ Space$. This toroidal structure optimizes QTV by leveraging quantum coherence if $\psi_0$ and $\psi_1$ are separable, non-entangled, tensor product states of pure states $\phi_0, \chi_0$ from $\mathcal{H}_A$ and $\phi_1, \chi_1$ from $\mathcal{H}_B$ as follows:*

$$|\psi_0\rangle = |\phi_0\rangle \otimes |\chi_0\rangle, \quad |\psi_1\rangle = |\phi_1\rangle \otimes |\chi_1\rangle \tag{29}$$

**Proof.** Let us track the two phase variables:

$$\theta(t) = E_0 t, \qquad \phi(t) = E_1 t.$$

These angles define independent circular motions on $S^1$, i.e., a 1-dimensional manifold (unit circle). The global state $\psi(t)$ thus depends on two angular variables $(\theta, \phi)$, which evolve independently. Therefore, the family of states $\psi(t)$ (up to a global phase) lies on the 2-dimensional torus:

$$\psi(t) \in \mathbb{T}^2 = S^1_\theta \times S^1_\phi. \tag{30}$$

The geometry of the state space becomes a torus when:

- The system is restricted to a 2D subspace spanned by $\psi_0, \psi_1$, or 2-torus embedded in the Hilbert space defined by Equation (29).
- The energies $E_0 \neq E_1$ generate two independent phase rotations,
- A quantum state as a superposition of two eigen states with distinct energies, and evolves coherently in time.
- The manifold represents continuous superposability of geometric positions or the smooth additive structure that makes calculus, flows, and trajectories possible.
- The tensor product $\otimes$ represents structural composition, not superposition. It combines independent degrees of freedom without interference.

Consider the Gaussian curvature metric of the torus

$$ds^2 = d\theta^2 + d\phi^2 = 0, \quad (\theta, \phi) \in [0, 2\pi)^2. \tag{31}$$

The Gaussian curvature is zero everywhere. This flat torus is locally indistinguishable from a plane. QTV is maximized when $\psi(t)$ aligns the best to the top eigen state of the measurement $H$ as shown in Equation (24). ∎

The torus structure emerges from the phase evolution under coherent superposition. The quantum interference patterns that define measurement coherence (e.g., striped amplitude patterns in Fig. 4) are directly encoded in the geometry of the torus. Both state and measurement must be quantum to manifest interference and non-classical behavior as shown in Equation (24) and (27). This result generalizes a superposition of $n$ orthogonal eigenstates with distinct energies $\{E_0, E_1, \ldots, E_{n-1}\}$, the system evolves on an $n$-dimensional torus:

$$\mathbb{T}^n = S^1 \times S^1 \times \cdots \times S^1. \tag{32}$$

The torus is an *emergent geometry* of phase evolution under coherent superposition. Each distinct energy eigenvalue $E_k$ induces a phase evolution $e^{-iE_k t}$, corresponding to a rotation on a manifold $S^1$. QTV is maximized when $\psi(t)$ aligns the best to the top eigen state of the measurement $H$.

## IV. Algorithm 1: Updating $H(t)$ Using a Wavefunction $\psi(t)$ from an Entangled Bipartite System

In this section, an algorithm computes and evolves Hamiltonian $H(t)$ based on the Page's Theorem. Take an initially pure, separable state in space $\mathcal{H}_A$ such as a torus, embed it into a larger composite space $\mathcal{H}_A \otimes \mathcal{H}_B$, and then apply a unitary rotation that mixes those subspaces.

We have the following algorithm:

### 1. Initialization

Start with a structured, asymmetric, nonnegative matrix $H_0 \in \mathbb{R}^{n \times n}$ that represents a predefined feedforward and feedback dynamics. Initialize:

$$H(0) = H_0$$

### 2. Generating an Entangled State

Let $\mathcal{H}_B \cong \mathbb{C}^n$, $\mathcal{H}_A \cong \mathbb{C}^d$ be two Hilbert spaces corresponding to subsystems $B$ and $A$. Their joint system is described by the tensor product space:

$$\mathcal{H}_{BA} = \mathcal{H}_B \otimes \mathcal{H}_A$$

We generate a random pure state $|\psi\rangle_{BA} \in \mathcal{H}_{BA}$ by drawing coefficients $c_{ij} \in \mathbb{C}$ from a complex Gaussian distribution and normalizing:

$$|\psi\rangle_{BA} = \sum_{i=1}^{n} \sum_{j=1}^{d} c_{ij} |i\rangle_B \otimes |j\rangle_A \tag{33}$$

where $\sum_{i,j} |c_{ij}|^2 = 1$

This results in a generic entangled state unless $|\psi\rangle_{BA}$ factorizes as $|\phi\rangle_B \otimes |\chi\rangle_A$. The density matrix of the total system is:

$$\rho_{BA} = |\psi\rangle_{BA} \langle\psi|_{BA} \tag{34}$$

The reduced density matrix for subsystem $B$ is obtained by tracing out subsystem $A$:

$$\rho_B = Tr_A(\rho_{BA}) = \sum_{j=1}^{d} \langle j|_A \, \rho_{BA} \, |j\rangle_A \quad (35)$$

This gives a mixed state $\rho_B \in \mathbb{C}^{n \times n}$, which is generally not pure due to entanglement with $A$.

*3. Evolving Rule for $H(t)$*

Let $H \in \mathbb{C}^n \times \mathbb{C}^n$ and $\psi \in \mathbb{C}^n \otimes \mathbb{C}^d$ be a pure bipartite quantum state, at each time step $t$, evolve $H(t)$ via a structured, masked, rank-1 additive deformation:

$$H(t+1) = (1-\lambda)H(t) + \eta \cdot \left(\psi(t)\psi(t)^\dagger \circ K\right) \quad (36)$$

where:
- $\eta > 0$ is the learning rate,
- $K \in \{0,1\}^{n \times n}$ is a binary mask defined by $K_{ij} = 1$ if $H_0[i,j] > 0$, and 0 otherwise,
- $\circ$ denotes elementwise (Hadamard) multiplication.

This update reinforces only the allowed (nonzero) pathways in $H_0$, keeping the matrix asymmetric and sparse. Equation (36) is more generic form than Equation (6).

## V. Use Case: Real-Life Quantum Process with Unknown or Emergent $H$ - Kill Web

In a realistic quantum process applied to a typical F2T2EA application, the measurement observable $H$ can be considered as a blue kill web of associations between entities of C2 plan & authorities, sensors, platforms, and weapons as shown in Fig. 5. $H$ is determined and optimized based on various factors:
- Red state of adversaries
- The environment such as weather or terrain

Therefore, we model blue's kill web as the measurement $H$ acts on the quantum red state $|\psi_{red}\rangle$ to make benefits blue as much as possible, that is to maximize the norm of $H\psi_{red}$ or the measurement of $\psi_{red}$ on $H^\dagger H$. $H$ can be initiated from a traditional kill chain model. For example, considering the interaction feasibility among each element in a kill chain, a matrix $\in \mathbb{R}^{10 \times 10}$ encodes a directional dependency structure among four categories of capabilities, each category of capabilities can be viewed as a CSOA:
- C2: index 0
- Sensor: indices 1–2 (dependent on C2)
- Platform: indices 3–5 (dependent on Sensor)
- Weapon: indices 6–9 (dependent on Platform and partially recurrent)

A kill web feasiblity matrix $K$ can be specified as in Equation (37):

$$K = \begin{bmatrix} 1 & 0 & 0 & 0 & 0 & 0 & 0 & 0 & 0 & 0 \\ 1 & 1 & 0 & 0 & 0 & 0 & 0 & 0 & 0 & 0 \\ 1 & 0 & 1 & 0 & 0 & 0 & 0 & 0 & 0 & 0 \\ 0 & 1 & 1 & 1 & 0 & 0 & 0 & 0 & 0 & 0 \\ 0 & 1 & 1 & 0 & 1 & 0 & 0 & 0 & 0 & 0 \\ 0 & 1 & 1 & 0 & 0 & 1 & 0 & 0 & 0 & 0 \\ 0 & 0 & 0 & 1 & 1 & 1 & 1 & 0 & 0 & 0 \\ 0 & 0 & 0 & 1 & 1 & 1 & 0 & 1 & 0 & 0 \\ 0 & 0 & 0 & 1 & 1 & 1 & 0 & 0 & 1 & 0 \\ 0 & 0 & 0 & 1 & 1 & 1 & 0 & 0 & 0 & 1 \end{bmatrix}, \quad (37)$$

where "1"s represents a feasible interaction between two components in the kill web, i.e., probabilities of usage of two capabilities in the same time or sequentially, among the different categories of capabilities.

With a quantum configuration, we can assume it is generated from a wave function that uses a superposition of C2 (C), Sensor (S), Platform (P), and Weapons (W) as in Equation (38):

$$|\psi_{\text{blue}}(t)\rangle = \sqrt{\lambda_0}e^{-iE_0t}|C\rangle + \sqrt{\lambda_1}e^{-iE_1t}|S\rangle + \sqrt{\lambda_2}e^{-iE_2t}|P\rangle + \sqrt{\lambda_3}e^{-iE_3t}|W\rangle, \quad (38)$$

where
- $|C\rangle, |S\rangle, |P\rangle, |W\rangle \in \mathbb{C}^n$ are fixed orthonormal basis vectors,
- $\lambda_j \geq 0$ and $\sum_{j=0}^{3} \lambda_j = 1$ are probability weights, $n_C, n_S, n_P,$ and $n_W$ are number of options in each category $C, S, P,$ and $W$ ($n = n_C + n_S + n_P + n_W$).
- $E_j \in \mathbb{R}$ control the time evolution phase.
- To customize Step 2 in Algorithm 1, we
  - Start with a separable state: The initial quantum state of the joint system $BA$ is a tensor product of two independent states since $\psi_B = \psi_{\text{blue}}(t)$ is a rank-1 wave function or a pure state:
    $$|\psi\rangle_{BA} = |\psi_B\rangle \otimes |\phi_A\rangle$$
    This state is *separable*.
  - Apply a global unitary operator: A unitary transformation $U_{BA}$ is applied to the joint system:
    $$|\psi'\rangle_{BA} = U_{BA}(|\psi_B\rangle \otimes |\phi_A\rangle) \quad (39)$$
    In general, $|\psi'\rangle_{BA}$ is no longer separable and may be *entangled*, depending on the structure of $U_{BA}$. Tracing out subsystem $A$ will then yield a mixed state on $B$, indicating entanglement.

This state evolves in time according to phase rotations and does not require the vectors $|C\rangle, \ldots, |W\rangle$ to be eigenvectors of the Hamiltonian $H$.

$|\psi_{\text{blue}}(t)\rangle$ can act as a low-rank approximation or become the top eigenvector of a evolved $H$:

$$H(t+1) = (1-\lambda)H(t) + \eta \left[\psi_{blue}(t)\psi_{blue}(t)^\dagger\right] \circ K \quad (40)$$

This is a QAET-QIS process and the the QTV is he measure of performance (MOP) of such a kill web against any adversarial input:
- $\psi_{\text{blue}}(t)$ lives on a multi-dimensional torus due to the phase rotations $e^{-iE_jt}$. It remains a superposition with toroidal structure in phase space.
- $\psi_{blue}(t)\psi_{blue}(t)^\dagger$ is a rank-1 positive semi-definite density matrix. The evolve of $H$ is coherent, structured, and constructive.
- By maximizing the spectral norm $\|H_t\| = \lambda_{\max}(H_t)$ over the parameters $\lambda_j$ and $E_j$, we can align $\psi_{blue}(t)\psi_{blue}(t)^\dagger$ with the dominant eigenvector of $H_t$.

The final $H$ is shown in Equation (41).

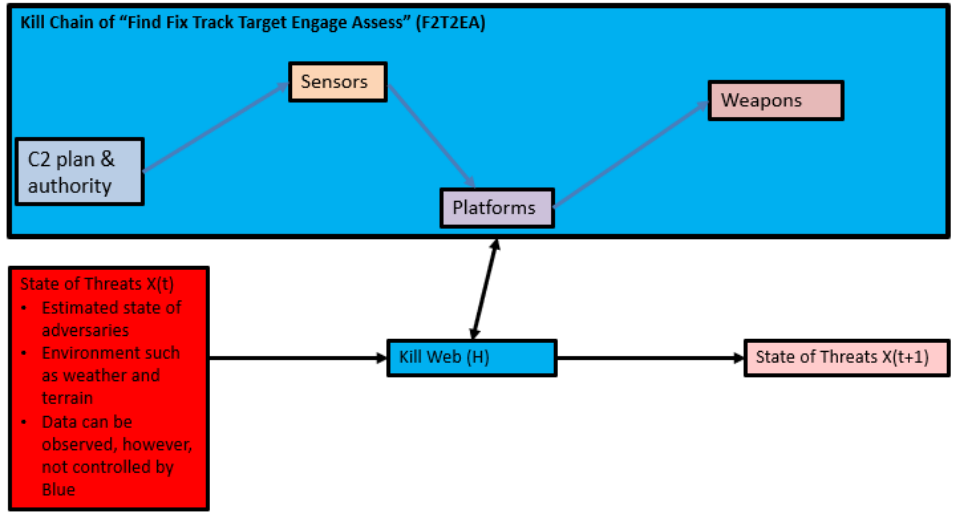

Fig. 5: Kill Web Concept

Fig. 6 shows the evolution of $H$ in terms of heatmaps, when $T = 500$, the best with the spectral norm of 0.605 (t=226) and the last with spectral norm of 0.427 (t=499). Fig. 7 shows the trend of spectral norm in the evolution. The trace of $H$ is the same across all the comparison.

$$\begin{bmatrix} 0.05 & 0 & 0 & 0 & 0 & 0 & 0 & 0 & 0 & 0 \\ 0.06 & 0.08 & 0 & 0 & 0 & 0 & 0 & 0 & 0 & 0 \\ 0.03 & 0 & 0.02 & 0 & 0 & 0 & 0 & 0 & 0 & 0 \\ 0 & 0.19 & 0.10 & 0.45 & 0 & 0 & 0 & 0 & 0 & 0 \\ 0 & 0.06 & 0.03 & 0 & 0.08 & 0 & 0 & 0 & 0 & 0 \\ 0 & 0.05 & 0.02 & 0 & 0 & 0.03 & 0 & 0 & 0 & 0 \\ 0 & 0 & 0 & 0.13 & 0.05 & 0.04 & 0.04 & 0 & 0 & 0 \\ 0 & 0 & 0 & 0.19 & 0.07 & 0.05 & 0 & 0.08 & 0 & 0 \\ 0 & 0 & 0 & 0.17 & 0.06 & 0.04 & 0 & 0 & 0.07 & 0 \\ 0 & 0 & 0 & 0.19 & 0.06 & 0.04 & 0 & 0 & 0 & 0.08 \end{bmatrix}$$
(41)

*A. Discussion*

1) The QAET-QIS framework ensures a kill web's performance against any adversarial input is optimized. The final optimized $H$ has the characteristics of global optimization, distributed lethality, and load balance.
2) The quantum superposition on a torus formed by different CSOAs in Equation (38), the QAET-QIS evolving rule for measurement $H$ in Equation (36) and Equation (40), are necessary to maximize the spectral norm of $H$, consequently maximize the kill web performance against any adversarial input.
3) Theorem 1 and 2, Equation (16), Equation (40), and Algorithm 1 in Section IV ensure the kill web $H$ has a maximum bound value 1. In other words, if each element $H$ is a probability of various capabilities used together or sequentially in a kill web $H$, also with the constraint of trace 1, the performance for $H$ against any adversarial input is bounded by 1, however, can be optimized using Algorithm 1.
4) Since the feasibility filter $K$ is asymmetrical, and possible entangled formation $H$ due to the selection of sub space in Equation (39), the maximum value 1 may not be achieved, only 0.605 in the example.
5) The entanglement allows B (blue) learn and exhibit coordinated dynamics from environment and adversary (A). Without it, B is perfectly coherent but isolated. The QAET-QIG mechanism in Equation (40) maintains the balance: structured internal coherence with controlled, beneficial entanglement for CSOAs of $\psi_{blue}(t)$ to be adaptive, self-organizing and context-sensitive.

## VI. CONCLUSION

In this paper, we show a QAET-QIG framework, where a repeated measurement, i.e., updating $H$ forced by the environment with CSOAs in a QAET process. We apply the QAET-QIG framework to the kill web concept that potentially improve the traditional kill chain process such as find, fix, track, target, engage, and assess (F2T2EA). The framework ensures the kill web performance against any adversarial input is optimized. The improvement is measured in the values of global optimization, distributed lethality, and load balancing.

## ACKNOWLEDGMENT

The authors would also like to thank the collaborative research of Quantum Intelligence, Inc. The views and conclusions contained in this document are those of the author and should not be interpreted as representing the official policies, either expressed or implied of the U.S. Government.

## REFERENCES

[1] Zhou, C., Zhao, Y., & Kotak, C. (2009). The Collaborative Learning agent (CLA) in Trident Warrior 08 exercise. In Proceedings of the International Conference on Knowledge Discovery and Information Retrieval - Volume 1: KDIR, IC3K (pp. 323-328) DOI: 10.5220/0002332903230328. Madeira, Portugal.
[2] Zhao, Y., & Zhou, C. (2023). Quantum Theoretic Values of Collaborative and Self-organizing Agents. ASONAM '23: Proceedings of the 2023 IEEE/ACM International Conference on Advances in Social Networks Analysis and Mining November 2023. Pages 678-685. Kusadasi, Turkiye, November 6 - 9, 2023. https://dl.acm.org/doi/10.1145/3625007.3627509
[3] Zhou, C. C. & Zhao, Y. (2023). Crowd-Sourcing High-Value Information via Quantum Intelligence Game. In: Arai, K. (eds) Intelligent Computing. SAI 2023. Lecture Notes in Networks and Systems, vol 711. Springer, Cham. https://doi.org/10.1007/978-3-031-37717-4_34.

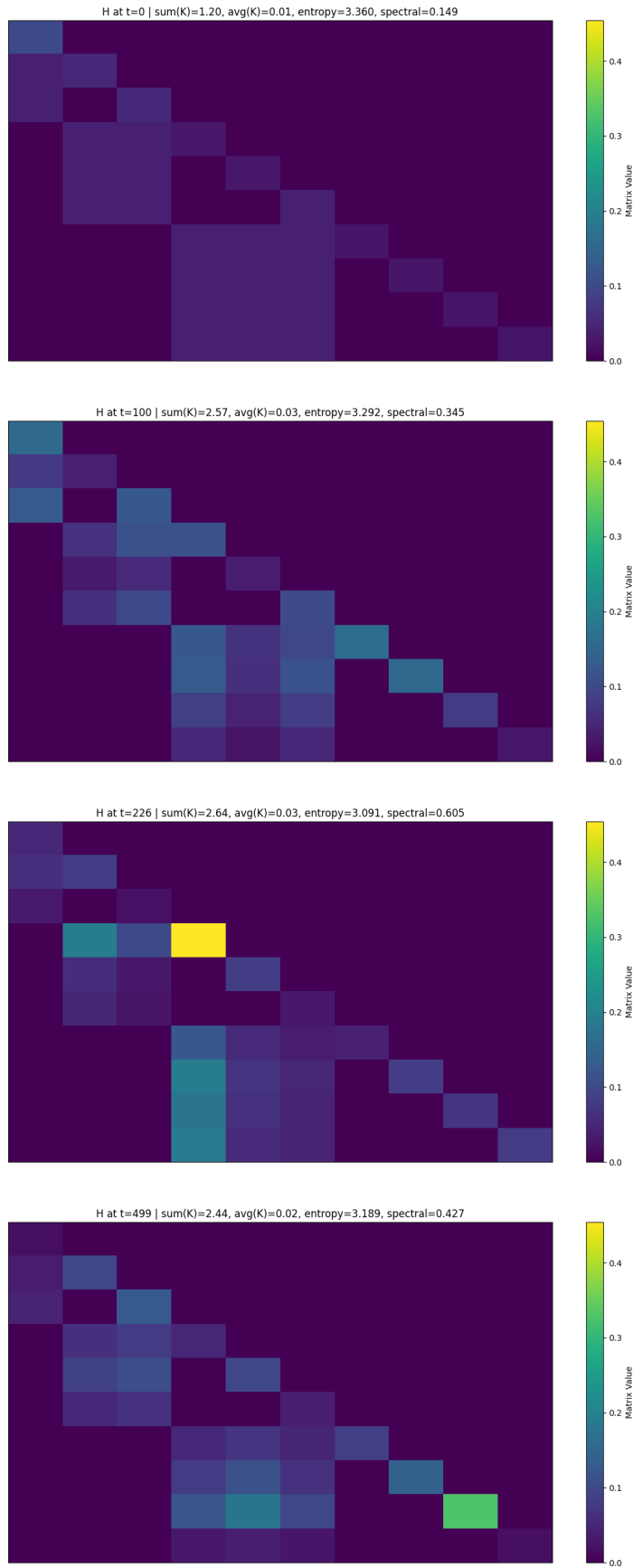

Fig. 6: Kill Web Heatmap Evolutions

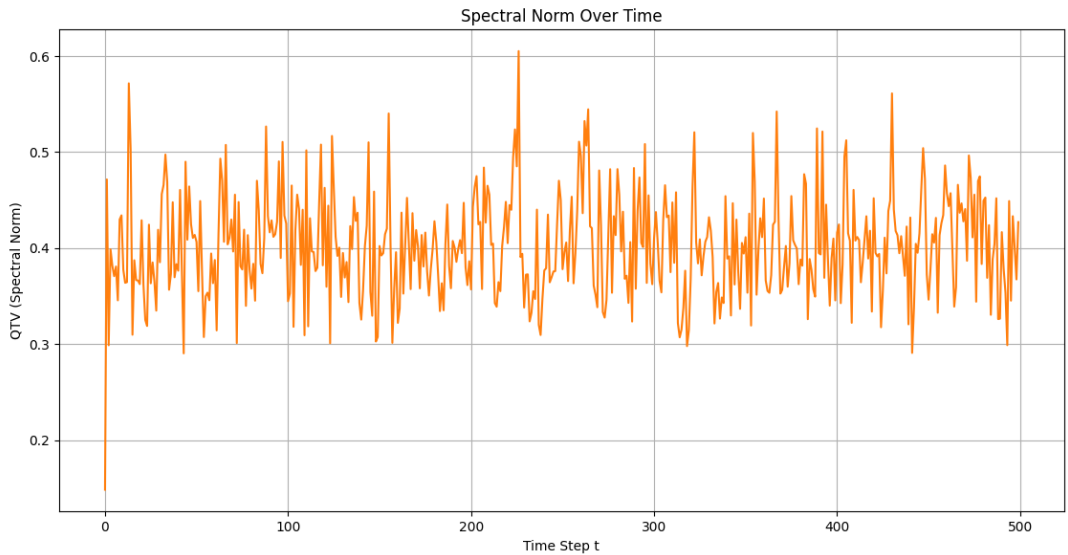

Fig. 7: $\eta = 0.7, \lambda = 0.7, d = 1$. The Trend of Spectral Norm in the Evolution.

[4] Zhao, Y., Mata, G. & Zhou, C. (2023). Self-organizing and Load-Balancing via Quantum Intelligence Game for Peer-to-Peer Collaborative Learning Agents and Flexible Organizational Structures. In: Arai, K. (eds) Intelligent Computing. SAI 2023. Lecture Notes in Networks and Systems, vol 711. Springer, Cham. https://link.springer.com/chapter/10.1007/978-3-031-37717-4_33

[5] Wen, Xiao-Gang (1990). Topological Orders in Rigid States. Int. J. Mod. Phys. B. 4 (2): 239. doi:10.1142/S0217979290000139.

[6] Farhi, E., Goldstone, J., & Gutmann, S. (2014). A Quantum Approximate Optimization Algorithm, arXiv Prepr. arXiv1411.4028, pages 1-16, 2014.

[7] S. Dusuel, M. Kamfor, R. Orus, K. P. Schmidt, J. Vidal (2011). Robustness of a perturbed topological phase. https://arxiv.org/abs/1012.1740

[8] Zhao, Y., and Zhou, C. (2025). Scoring the Impact of Unstructured Data Using Quantum Properties. In the International Symposium on Foundations and Applications of Big Data Analytics (FAB 2025), Aug 24- 28, 2025, Niagara Falls, Ontario, Canada

[9] Yingwen Zhang, Dao-Xin Yao, and Zhi Wang (2023). Topological superconductivity with large Chern numbers in a ferromagnetic-metal–superconductor heterostructure. Phys. Rev. B 108, 224513 – Published 18 December 2023

[10] John von Neumann, J. (1955). Mathematical Foundations of Quantum Mechanics. Princeton University Press. ISBN 978-0-691-02893-4.

[11] Barile, M. & and Weisstein, E. W. (2023). Betti Number." https://mathworld.wolfram.com/BettiNumber.html

[12] Che, Y.,Gneiting, C., T Liu, T. & Nori, F. (2020). Topological quantum phase transitions retrieved through unsupervised machine learning. Physical Review B, 2020.https://arxiv.org/pdf/2002.02363.pdf

[13] Continentino, M. (2017). Quantum Scaling in multi-agent Systems – An Approach to Quantum Phase Transitions. Cambridge University Press.

[14] Kartik, Y. R. Kumar, R.R. & Sarkar, S. (2020). Topological Quantum Phase Transitions and Criticality in a Longer-Range Interacting Kitaev Chain. Physical review B 9 September 2020.

[15] Hofstadter, D. (2004). Teuscher, C. (ed.). Alan Turing: Life and Legacy of a Great Thinker. Springer. p. 54. ISBN 978-3-540-20020-8.

[16] Meyer C. D.(2000), Matrix Analysis and Applied Linear Algebra. SIAM, ISBN 978-0-89871-454-8

[17] Newman, M. E. J. (2006). Finding community structure in networks using the eigenvectors of matrices. Phys. Rev. E 74, 036104.

[18] Lotidis, K., Mertikopoulos, P., and Bambos, N. (2023). Learning in Quantum Games.

[19] Zhao, Y. (2024). Quantum Theoretic Values of Collaborative and Self-Organizing Agents in Forming a Hybrid Force. In the 29th ICCRTS Proceedings, 24-26 September 2024 · London, UK.

[20] Page, D. N.(1993). Average entropy of a subsystem. Physical Review Letters, 71(9), 1291–1294 (1993).DOI: 10.1103/PhysRevLett.71.1291

[21] Farhi, E., Goldstone, J.,Gutmann, S., & Sipser, M. (2000). Quantum Computation by Adiabatic Evolution. MIT CTP # 2936 quant-ph/000110